\documentclass[twocolumn,showpacs,preprintnumbers,amsmath,amssymb]{revtex4}
\usepackage{amsfonts}
\usepackage{amsmath}
\usepackage{bbm}
\usepackage{txfonts}

\usepackage{graphicx}
\usepackage{dcolumn}
\usepackage{bm}

\begin{document}

\preprint{APS/123-QED}

\title{Quantum repeaters based on Rydberg-blockade coupled atomic ensembles}

\author{Yang Han$^{1,2}$, Bing He$^1$, Khabat Heshami$^1$, Cheng-Zu Li$^2$, and Christoph Simon$^1$}
\affiliation{$^1$ Institute for Quantum Information Science
and Department of Physics and Astronomy, University of
Calgary, Calgary T2N 1N4, Alberta, Canada\\$^2$ College of Science, National University of Defense
Technology, Changsha 410073, China.}

\date{\today}

\begin{abstract}
We propose a scheme
for realizing quantum repeaters with Rydberg-blockade coupled
atomic ensembles, based on a recently proposed collective encoding strategy. Rydberg-blockade mediated two-qubit gates and
efficient cooperative photon emission are employed to create
ensemble-photon entanglement. Thanks to deterministic entanglement swapping operations via Rydberg-based two-qubit gates, and to the suppression of multi-excitation errors by the blockade effect, the entanglement distribution rate of the present scheme is higher by orders of magnitude than the rates achieved by other ensemble-based repeaters. We also show how to realize temporal multiplexing with this system, which offers an additional speedup in entanglement
distribution.
\end{abstract}

\pacs{03. 67.Hk, 32.80.Ee, 42.50.Ex}
\maketitle
\section{\label{sec:level1}INTRODUCTION }

Quantum communication aims at secure message transmission
between remote locations by employing entanglement for
 quantum teleportation \cite{1} or quantum cryptography
 \cite{2}. Unfortunately, since the inevitable photon loss scales exponentially
 with the length of channel, it is difficult to establish
 high quality entanglement over long distances. This problem
 may be overcome by quantum repeaters \cite{3}, which create
 and store shorter-distance entanglement in a heralded way,
 and then connect the elementary entangled states to establish longer-distance entanglement
 via entanglement swapping.

A highly influential protocol for realizing quantum repeaters was
proposed by Duan, Lukin, Cirac and Zoller (DLCZ)\cite{4}. It is
based on macroscopic atomic ensemble quantum memories, Raman
scattering and linear optics. There is a significant body of
theoretical \cite{5,6,7,8} and experimental \cite{9} work based on this general approach. In this paper, we refer to these
schemes as DLCZ-type repeaters. The significant advantage of
DLCZ-type repeaters is that they use relatively simple elements.
However, there are two intrinsic limitations in this approach.
First, as Raman scattering is used to create entanglement between a single atomic excitation and a single photon, inevitable multi-excitation (and multi-photon) terms cause errors in the final states. In order to suppress multiple excitations, one has to work with very low excitation probability \cite{4}. Second, the Bell measurements in the swapping
operations are realized via linear optics, so the success
probability of the entanglement swapping is bounded by 1/2
\cite{10}. The above two limitations significantly diminish the
efficiency of DLCZ-type repeaters.

There are a number of proposals for realizing quantum
repeaters using ingredients other than atomic ensembles and
linear optics \cite{11,12,13}. Most of them use individual
quantum systems as the quantum memory \cite{12,13}. An
obvious advantage of using individual quantum systems is
that the problem of multiple excitations is eliminated. If
the two-qubit gates for Bell measurement in the swapping
operations can also be realized efficiently, repeaters
based on individual quantum systems have the potential to
significantly outperform DLCZ-type repeaters \cite{12,13}.
However, for the individual quantum systems one has to
precisely address every single particle, and one may need
cavities to achieve a high efficiency of photon collection
\cite{13}.

An attractive technique for quantum information processing (QIP)
with atomic ensembles is based on the Rydberg blockade mechanism, cf. below. There have been a number of proposals to use the Rydberg blockade for various QIP tasks(see \cite{14} for an overview). In
the present paper, we propose a concrete scheme for realizing quantum repeaters in this way and analyze its performance in detail. We show that the entanglement distribution rate offered by repeaters based on Rydberg blockade coupled ensembles significantly surpasses the rate of DLCZ-type repeaters. Compared to the schemes involving individual quantum systems, repeaters based on Rydberg blockade coupled ensembles achieve almost the same distribution rate and avoid addressing single particles and using cavities.
Our proposed scheme also allows temporal multiplexing \cite{15},
which could further enhance the achievable distribution rate.

\section{Rydberg blockade coupled atomic ensemble}

Rydberg states are states of alkali atoms characterized by
a high principal quantum number. Atoms in such Rydberg
states have large size and can therefore have large dipole
moments, resulting in strong dipole-dipole interactions
\cite{16}. Due to this strong long-range interaction, a
single atom in an atomic ensemble excited to a Rydberg
state shifts the Rydberg energy level of its neighbors out
of resonance and blocks further excitations, which is
called the Rydberg blockade mechanism \cite{17}, and this
kind of ensembles is referred to as Rydberg blockade
coupled ensembles \cite{14}. Recently, experiments have
demonstrated an almost perfect blockade \cite {18} as well
as a blockade-based C-NOT gate \cite {19} between a single
pair of trapped atoms at separation $R\leq10\mu m$.
Although no experiments have been done with an ensemble
where the blockade acts across the whole ensemble, a number
of experiments show clear signs of the blockade effect on
larger samples \cite{14}.

In this blockade regime, an effective two-level system is
realized between the state with all atoms in the ground
level and the single-excitation symmetric atomic state.
This two-level system has an effective light-atom coupling
that is a factor of $\sqrt{N}$ larger than the
light-single-atom coupling. It is promising for a wide
variety of quantum information processing applications
\cite{20,21,22,23,24,25,26}. In the following two
subsections, we briefly review the collective encoding
strategy for a \emph{k}-bit quantum register \cite{23} and
the cooperative photon emission effect \cite{25,26,27,28}
in a Rydberg blockade coupled ensemble, which are directly
related to our repeater scheme.

\subsection{Collective encoding in a Rydberg blockade coupled ensemble}

A Rydberg coupled atomic ensemble consisting of \emph{ N}
atoms can be used to build a \emph{k}-bit quantum register
$(N\gg k)$, where the qubits are collectively encoded in
different single excitation symmetric atomic states
\cite{23}. As shown in Fig.1(a), \emph{k} qubits are
encoded in an ensemble of \emph{N} atoms with $2k+1$
long-lived ground levels.  The $i$-th qubit values zero and
one are identified with symmetric single-excitation atomic
states populating $|0_{i}\rangle$ and $|1_{i}\rangle$,
respectively. The initialization of this \emph{k}-bit
register is as follows: Originally all the atoms are in the
reservoir state $|g\rangle$. Then due to the blockade
mechanism, one can transfer precisely one atom to each pair
of levels $(|0_{i}\rangle,|1_{i}\rangle)$ via a Raman
transition involving a Rydberg state $|r\rangle$.

In the following, for given atomic levels $|x_{i}\rangle
(x=0,1, i=1\cdots k)$, we will let the kets
$|\tilde{x_{i}}\rangle (x=0,1, i=1\cdots k)$ denote the
symmetric single-excitation collective atomic states, for
instance,
\begin{equation}
|\tilde{1}_{1}\rangle=\frac{1}{\sqrt{N}}\sum_{j=1}^{N}e^{-i\vec{k}_{0}\cdot
\vec{r}_{j}}|g\rangle_{1}|g\rangle_{2}\cdots|1_{1}\rangle_{j}\cdots|g\rangle_{N},
\end{equation}
where $\vec{r}_{j}$ is the position of the $j$-th atom,
$|1_{1}\rangle_{j}$ indicates that the $j$-th atom is in
the state $|1_{1}\rangle$, and $\vec{k}_{0}$ is the
summation of all the wave vectors of the light pulses used
to transfer the ensemble to the above state, i.e.,
$\vec{k}_{0}=\sum_{m}\lambda_{m} \vec{k}_{m}$, and
$\lambda_{m}=\pm1$, depending on whether a photon is
absorbed or emitted during the $m$-th pulse. Accordingly,
the basis of the $i$-th qubit can be written as
$(|\tilde{0_{i}}\rangle,|\tilde{1_{i}}\rangle)$.

It has been proposed in Ref. \cite{23} that both single-bit
rotations and two-qubit gates can be realized in this
system. The single-qubit rotations on the $i$-th qubit are
straightforwardly performed using two-photon stimulated
Raman beams coupling $|0_{i}\rangle$ and $|1_{i}\rangle$.
The two-qubit phase gate between the $j$-th and $k$-th
qubits is implemented by a sequence of three laser pulses
as shown in fig.1(b): (\emph{i}) The excitation of the
control qubit internal state $|0_{j}\rangle$ into one
Rydberg state $|r_{1}\rangle$; (\emph{ii}) $2\pi$ Rabi
rotation between the target qubit state $|0_{k}\rangle$ and
another Rydberg state $|r_{2}\rangle$; (\emph{iii}) The
return of the population from $|r_{1}\rangle$ to
$|0_{j}\rangle$. If the control qubit is in state
$|0\rangle$, the resulting unit occupancy of the
$|r_{1}\rangle$ state blocks the Rabi cycle and nothing
happens to the target qubit, while a control qubit in state
$|1\rangle$ causes no blockade, and hence we obtain a
controlled $\pi$ phase shift on the $|0_{k}\rangle$ state
amplitude due to a full Rabi cycle. Hence, universal
quantum computing operations can be realized in this system
\cite{23}.

\begin{figure}
\scalebox{0.8}{\includegraphics*[23,644][299,825]{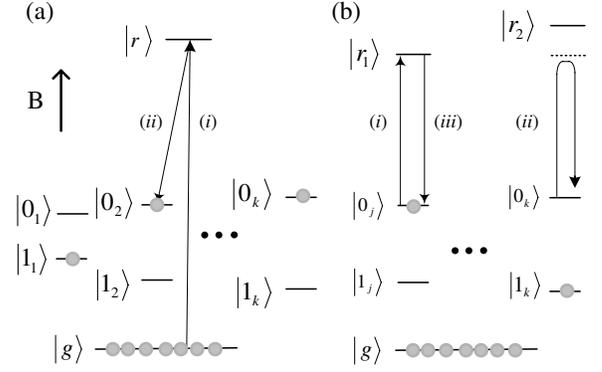}} 
\caption{\label{fig:epsart} (a) Collective encoding of \emph{k}
qubits in an ensemble of \emph{N} atoms (gray dots) with $2k+1$
long-lived levels; $|r\rangle$ is a Rydberg state, and
$|x_{i}\rangle(x=0,1  i=1\cdots k)$ and $|g\rangle$ are ground
states. A magnetic field $B$ is applied to the ensemble, which
enables every ground state to be selectively manipulated via
appropriate choices of laser frequencies. The transitions (\emph{i})
and (\emph{ii}) show the initialization procedure of the second
qubit. (b) Two-qubit phase gate on collectively encoded qubits $j$
and $k$ mediated by Rydberg blockade. See text for details.}
\end{figure}

\subsection{Cooperative photon emission from a Rydberg blockade coupled ensemble}
Now we discuss how to map a collectively encoded qubit in
the ensemble into a flying photonic qubits. Assume the
ensemble is in the state $|\tilde{1_{f}}\rangle$. If one
transfers state $|\tilde{1_{f}}\rangle$ into an excited
state $|\tilde{e}\rangle$ via a $\pi$-pulse laser with wave
vector $\vec{k}_{e}$, this state will radiate into a
variety of modes with all the atoms in state $|g\rangle$
and a single-photon propagating with the wave vector
$\vec{k}$. The amplitude for emitting a photon with wave
vector $\vec{k}$ and polarization $\vec{e}$, is
proportional to
\begin{equation}
\langle g_{1}\cdots g_{N}|\langle \vec{k}|(\vec{e}\cdot
d)\hat{a}^{\dagger}_{k}|\tilde{e}\rangle|vac\rangle=-\sum_{j=1}^{N}\frac{\langle
g|(\vec{e}\cdot \vec{d})|e\rangle
e^{-i(\vec{k}_{0}+\vec{k}_{e}-\vec{k})\cdot \vec{r}_{j}}}{\sqrt{N}},
\end{equation}
where $\hat{a}^{\dagger}_{\vec{k}}$ is the creative operator for a
photon in mode $\vec{k}$ and $\vec{d}$ is the dipole operator. Thus
the transition probability $P(\vec{k})$ is proportional to
\begin{equation}
P(\vec{k})\propto\frac{1}{N}|\sum_{j=1}^{N}e^{-i(\vec{k}_{0}+\vec{k}_{e}-\vec{k})\cdot
r_{j}}|^{2}.
\end{equation}
Note that if $\vec{k}_{0}+\vec{k}_{e}=\vec{k}$ all the phase terms
are zero and $P(\vec{k})\propto N$; otherwise the phase terms become
random so that $P(\vec{k})\propto 1$, which means the emission is
highly directional. Although the typical size of a Rydberg blockade
coupled ensemble is less than 10$\mu m$ consisting of only several
hundreds of atoms \cite{14}, the above cooperative emission effect
is still large enough to ensure very good directed emissions of
photons \cite{25,26,27,28}, cf. below.

 There are two different ways to convert a collectively encoded qubit
$\alpha|\tilde{0_{f}}\rangle+\beta|\tilde{1_{f}}\rangle$
into a flying photonic polarization qubit
$\alpha|h\rangle+\beta|v\rangle$, where $h$ and $v$
indicate the horizontal and vertical polarization,
respectively. On the one hand, using polarization selection
rules, one can transfer the state $|\tilde{0}_{f}\rangle$
and $|\tilde{1}_{f}\rangle$ to excited states
$|\tilde{e}_{h}\rangle$ and $|\tilde{e}_{v}\rangle$,
respectively, and the subsequent atomic decay to the state
$|g_{1}\cdots g_{N}\rangle$ leads to emission of the
photonic state (Fig.2(a)). However, the horizontal and
vertical amplitudes may have different frequencies. While
this does not prevent the creation of entanglement between
remote ensembles, it makes it more challenging to prove
intermediate ensemble-photon entanglement experimentally.
On the other hand, one can use only one excited state
$|e\rangle$ and appropriately choose the wavevectors of the
lasers for transitions $(\vec{k}_{e}:|\tilde{0}_{f}\rangle\
\rightarrow|\tilde{e}\rangle)$ and
$(\vec{k}_{e}^{'}:|\tilde{1}_{f}\rangle\
\rightarrow|\tilde{e}\rangle)$. Thus the cooperative
emission from $|\tilde{e}\rangle$ to $|g_{1}\cdots
g_{N}\rangle$ is directed into different directions
$\vec{k}$ and $\vec{k}^{'}$ due to the phase matching
condition, as shown in Fig.2(b). Then the photonic state
$\alpha|\vec{k}\rangle+\beta|\vec{k}^{'}\rangle$ can be
easily changed into photonic polarization state
$\alpha|h\rangle+\beta|v\rangle$ by linear optics.

\begin{figure}
\scalebox{0.8}{\includegraphics*[14,649][291,827]{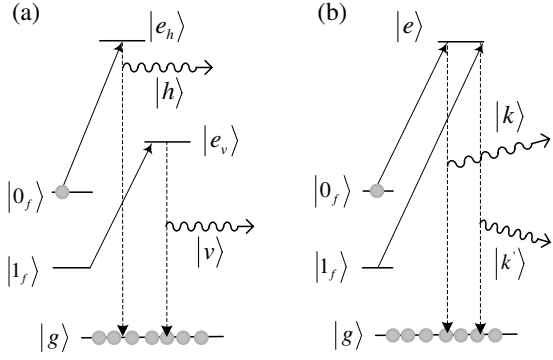}} 
\caption{\label{fig:epsart} Two different ways to convert a
collectively encoded qubit into a photonic qubit, cf.text.
}
\end{figure}

\section{repeater based on Rydberg blockade coupled ensembles}
\subsection{Main idea and efficiency}
In our scheme, each repeater node contains a single
ensemble collectively encoding three qubits. As shown in
Fig.3, qubit $s$ ($s$=1,2) is a stationary qubit, and qubit
$f$ is responsible for producing a flying photonic qubit.
To establish entanglement between ensemble $A$ and ensemble
$B$, we first focus on qubit 1 and qubit $f$ in these two
ensembles. In each ensemble, we prepare the entangled state
$(|\tilde{0_{1}}\rangle|\tilde{0_{f}}\rangle+|\tilde{1_{1}}\rangle|\tilde{1_{f}}\rangle)/\sqrt{2}$
using the single-bit and two-qubit gate described in the
previous section. Then qubit $f$ in each ensemble is
converted into a photonic polarization state via the method
shown in fig.2. The joint state of the two emitted photons
and of ensembles $A$ and $B$ can be expressed as
\begin{equation}
\begin{split}
&|\psi^{A}\rangle\otimes|\psi^{B}\rangle=\\
&(|\tilde{0_{1}}\rangle_{A}|h\rangle_{A}+|\tilde{1_{1}}\rangle_{A}|v\rangle_{A})\otimes
(|\tilde{0_{1}}\rangle_{B}|h\rangle_{B}+|\tilde{1_{1}}\rangle_{B}|v\rangle_{B})/2.
\end{split}
\end{equation}

\begin{figure}
\scalebox{0.9}{\includegraphics*[21,675][224,820]{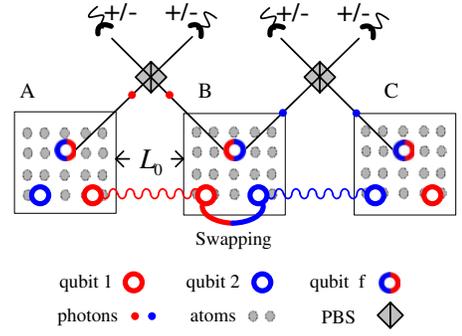}} 
\caption{\label{fig:epsart} (Color online) Quantum repeater
based on Rydberg blockade coupled ensembles. Neighboring
nodes are separated by a distance $L_{0}$. Each node
consists of one atomic ensemble encoding three qubits,
qubit 1 (red circle), qubit 2 (blue circle) and qubit f
(two-tone circle). Each ensemble asynchronously emits two
single photons (red and blue dots), which are entangled
with qubits 1 and 2 (red and blue circles), respectively.
Then the photons will be measured in the middle stations by
polarizing beam splitters (PBSs) and photon detectors. When
certain two-photon coincidences are observed, the 'red'
('blue') qubits in neighboring ensembles are projected into
entangled states. Then entanglement is extended to long
distances by entanglement swapping operations via two-qubit
gates on qubits 1 and 2 in the same ensemble.}
\end{figure}

Combining the two emitted photons on a polarizing beam
splitter (PBS) at a central station located half-way
between ensembles $A$ and $B$, a probabilistic Bell state
analysis can be performed by counting the photon number in
each output mode
$d_{\pm}=\frac{1}{\sqrt{2}}(|h\rangle_{A}\pm|v\rangle_{B})$
and
$\tilde{d}_{\pm}=\frac{1}{\sqrt{2}}(|h\rangle_{B}\pm|v\rangle_{A})$.
Such Bell analysis projects non-destructively the two
ensembles into an entangled state. In particular, the
detection of two photons, one in each mode $d_{+}$ and
$\tilde{d}_{+}$, leads to the entangled state
\begin{equation}
|\psi^{AB}\rangle=
(|\tilde{0_{1}}\rangle_{A}|\tilde{0_{1}}\rangle_{B}+|\tilde{1_{1}}\rangle_{A}|\tilde{1_{1}}\rangle_{B})/\sqrt{2}.
\end{equation}
In the ideal case, the probability for such an event is 1/8. Taking
into account the coincidences between $d_{+}-\tilde{d}_{-}$,
$d_{-}-\tilde{d}_{+}$ and $d_{-}-\tilde{d}_{-}$ combined with the
appropriate one-qubit operations, the probability to get the state
(5) is 1/2 (in the absence of transmission losses etc., cf. below).

In this manner, the entanglement can be established between
ensembles $A-B$, $C-D$, etc. To entangle the remaining
links, the procedure will be repeated with qubit 2 and
qubit f between ensembles $B-C$, $D-E$, etc. Considering
two links, say $A-B$ and $B-C$, the resulting state after
successful entanglement creation is

\begin{equation}
\begin{split}
&|\psi^{AB}\rangle\otimes|\psi^{BC}\rangle\\
&=(|\tilde{0_{1}}\rangle_{A}|\tilde{0_{1}}\rangle_{B}+|\tilde{1_{1}}\rangle_{A}|\tilde{1_{1}}\rangle_{B})(|\tilde{0_{2}}\rangle_{B}|\tilde{0_{2}}\rangle_{C}+|\tilde{1_{2}}\rangle_{B}|\tilde{1_{2}}\rangle_{C})/2.
\end{split}
\end{equation}

 We now calculate the time needed for
entanglement creation between two neighboring ensembles,
which are separated by a distance $L_0$. Let us denote by
$p$ the success probability for an ensemble to emit a
photon, including the probability to prepare the entangled
state between qubit $s$ and qubit $f$, the efficiency of
converting qubit $f$ into a photon and coupling it into the
fiber. The probability to get the expected twofold
coincidence is thus given by
$P_{0}=\frac{1}{2}p^{2}\eta^{2}_{d}\eta^{2}_{t}$, where
$\eta_{d}$ is the photon detection efficiency and
$\eta_{t}=exp\frac{-L_{0}}{2L_{att}}$ is the transmission
efficiency corresponding to a distance of
$\frac{L_{0}}{2}$, where $L_{att}$ is the fiber attenuation
length. Here we assume the losses in the fiber are
0.2dB/km, corresponding to $L_{att}=22km$. The entanglement
creation attempts can be repeated at time $t_{p}+t_{com}$,
where $t_{com}=L_{0}/c$ is the communication time and
$c=2\times10^{8} m/s$ is the light velocity in the fiber
\cite{15}. We assume a typical preparation $t_p =20 \mu$s,
cf. below. As a consequence, the average time required to
entangle two ensembles separated by a distance $L_{0}$ is
given by
\begin{equation}
T_{link}=\frac{t_{p}+t_{com}}{P_{0}}=\frac{2(t_{p}+L_{0}/c)}{p^{2}\eta^{2}_{d}\eta^{2}_{t}}.
\end{equation}

\begin{figure}
\scalebox{0.6}{\includegraphics*[83,266][490,570]{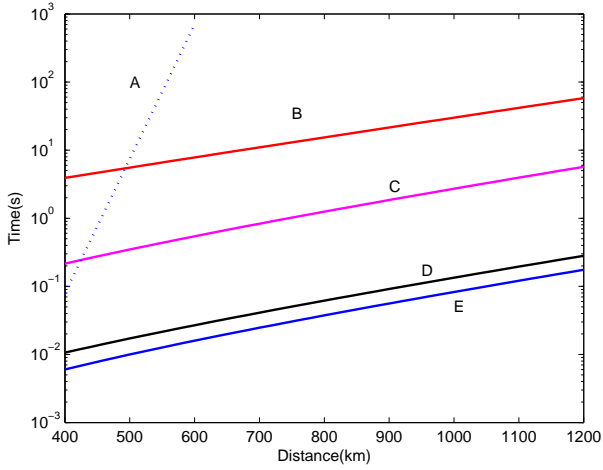}} 
\caption{\label{fig:epsart} (Color online) Performance of
quantum repeaters based on Rydberg blockade coupled
ensembles.  The quantity shown is the average time needed
to distribute a single entangled pair for the given
distance. We assume losses of 0.2 dB/km, corresponding to
telecom fibers at a wavelength of 1.5 $\mu$m. Curve A: as a
reference, the time required using direct transmission of
photons through optical fibers with a single-photon
generation rate of 10 GHz. Curve B: the most efficient
DLCZ-type repeater scheme known to us \cite{8}. In this
scheme high-fidelity entangled pairs are generated locally,
and entanglement generation and swapping operations are
based on two-photon detections. We have assumed memory and
detector efficiencies of 0.9. Curves C and D: schemes based
on Rydberg blockade coupled ensembles with $p=0.2$ and
$p=0.9$, respectively. Curve E: scheme based on trapped
ions in high-finesse cavities, where the success
probability for the ion to emit a photon is 0.9. Notice
that we imposed a maximum number of 16 links in the
repeater chain for curves B, C, D and E. }
\end{figure}

The entanglement can further be distributed over longer
distances by using successive entanglement swapping
operations between elementary links. Such swapping
operations require a local Bell state analysis, applied,
for example, on the qubit 1 and qubit 2 in ensemble $B$ to
entangle ensembles $A$ and $C$. Thanks to the high-fidelity
single-bit and two-qubit gates available in our system, the
success probability of entanglement swapping is only
restrained by the read-out efficiency. An effective
read-out mechanism in this context can be realized by
state-selective ionization \cite{29}. By coupling different
ground levels to different excited states and selectively
ionizing them, the resulting electron and ion can be
detected by channel electron multipliers. As it is
sufficient to detect at least one of the ionization
fragments, the overall detection efficiency can be as high
as $\eta_{ion}=95\%$ \cite{29}. Hence, the success
probability of entanglement swapping is
$P_{swap}=1/\eta_{ion}^{2}$, and the total time for the
distribution of an entangled pair over the distance
$2L_{0}$ is given by
\begin{equation}
T_{2L_{0}}=\frac{3}{2}\left(t_{p}+\frac{L_{0}}{c}\right)\frac{1}{P_{0}P_{swap}}=3\left(t_{p}+\frac{L_{0}}{c}\right)\frac{1}{
p^2  \eta_{t}^{2} \eta_{d}^{2}\eta_{ion}^{2}}.
\end{equation}
The factor 3/2 takes into account the fact that for the swapping
attempt one has to establish two neighboring links. If the
average waiting time for entanglement generation for one link is
$T$, there will be a success for one of the two after $T/2$ ; then
one still has to wait a time $T$ on average for the second one,
giving a total of $3T/2$. This simple argument gives exactly the
correct result in the limit of small $P_{0}$. In a
repeater with $n$ nesting levels, the precise values of analogous
factors have no analytic expression, but numerical results show
that this remains a good approximation \cite{30}. Hence, the total
time required for a successful entanglement distribution over the
distance $L=2^{n}L_{0}$ can be expressed as
\begin{eqnarray}
T_{tot}\approx
\left(\frac{3}{2P_{swap}}\right)^{n}\left(t_{p}+\frac{L_{0}}{c}\right)\frac{1}{P_{0}}\nonumber\\=\left(t_{p}+\frac{L_{0}}{c}\right)\frac{3^{n}}{2^{n-1}
p^2 \eta_{d}^{2} \eta_{t}^{2} \eta_{ion}^{2n}}.
\end{eqnarray}

We calculate the performance of a quantum repeater based on Rydberg
blockade coupled ensembles with Eq.(9), as shown in Fig. 4. In the
same figure we also show the performance of the most efficient
DLCZ-type repeater known to us \cite {8}, and  that of a repeater
based on trapped ions \cite {13}. One can see that the achievable
performance for repeaters based on Rydberg blockade coupled
ensembles greatly exceeds the best DLCZ-type repeater, and is
comparable with the repeater based on trapped ions with high-finesse
cavities. Another feature of repeaters based on Rydberg blockade
coupled ensembles is that the average time for the distribution of
an entangled pair scales only like $1/p^{2}$, in contrast to the
DLCZ-type repeaters which are much more sensitive to a reduction in
memory efficiencies. As can be seen from Fig.4, even with $p=0.2$,
the entanglement distribution time of our scheme is still 10 times
shorter than the time achievable with the best known DLCZ-type
repeater protocol.

\subsection{Implementation and noise analysis}

\begin{figure}
\scalebox{0.9}{\includegraphics*[13,650][191,831]{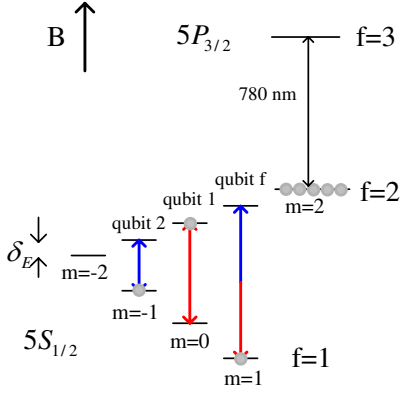}} 
\caption{\label{fig:epsart} (Color online) $^{87}$Rb level scheme
and identification of a three-qubit quantum register. See text for
details.}
\end{figure}

In this paper, we focus on realizing the present scheme
with $^{87}$Rb, whose nuclear spin $I=3/2$ provides 8
stable Zeeman states in the f=1,2 hyperfine levels (Fig.5).
A modest magnetic field of $B\sim 20G$ is applied to the
atoms, which gives splitting among the above Zeeman states
at least $\delta_{E}=\frac{\mu_{B}B}{2h}\sim 14$MHz, where
$\mu_{B}$ is the Bohr magneton. Initially, all the atoms
are in the reservoir state $|5S_{1/2}, f=2, m=2\rangle$.
Qubit s (s=1,2) and qubit f are encoded in other six Zeeman
states via collective encoding, as shown in Fig.5. The
state $|5P_{3/2},f=3,m=2\rangle$ is employed as $|e\rangle$
for mapping qubit f into photonic qubits so that the
wavelength of the emitted photons is $\lambda=780nm$.

In principle, any atomic ensembles that are suitable for collective
encoding strategy can be used to implement the present scheme
\cite{23}. To be specific, we first study a cubic lattice with
several hundreds of atoms. It should be noted that our scheme has a
significant flexibility of the shape and density of the ensemble, cf. below. For now let us suppose that an ordered three-dimensional array of $7\times7\times15=735$ $^{87}Rb$ atoms is loaded into an
elongated optical lattice. With a lattice spacing of $0.37 \mu m$,
the maximum distance between any two atoms is $R_{max}\approx 6 \mu
m$. We carefully choose two Rydberg states $|r_{1}\rangle$ and
$|r_{2}\rangle$, ensuring that the usual $C_{5}/R^{5}$ or
$C_{6}/R^{6}$ van-der-Waals interaction is resonantly enhanced by
F\"{o}ster processes leading to isotropic $C_{3}/R^{3}$ long-range
interaction \cite{19,22,23}. Assuming the principal quantum number
of the Rydberg states is $n=80$, a blockade shift as large as
$\textbf{B}/2\pi \geq10$MHz at a separation of $6 \mu m$ is
achievable \cite{19,23}.
\begin{table*}[!h]
\caption{The double excitation probability $P_{2}$,
spontaneous emission probability $P_{loss}$ and relevant
errors in four different procedures involving Rydberg
blockade in our scheme. We denote the Rabi frequencies
corresponding to qubit $s$ and qubit $f$ by $\Omega_{s}$
and $\Omega_{f}$, respectively; $\tau$ is the lifetime of
Rydberg state and $d$ is the optical depth of the ensemble.
Since qubit $s$ and qubit $f$ play different roles in
different procedures, the errors caused by $P_{2}$ and
$P_{loss}$ need to be calculated separately for each. See
text for details.}
\renewcommand{\arraystretch}{1.2}
\begin{center}
\begin{tabular}{llll}

\hline \hline

Procedures involving Rydberg blockade & $P_{2}$ & $P_{loss}$ & Error caused by $P_{2}$ and $P_{loss}$ \\[0.5ex] \hline
Initializing qubit s 　　　　&
$P^{(is)}_{2}=\Omega^{2}_{s}/2\textbf{B}^{2}$  &
$P^{(is)}_{loss}=\pi/\tau \Omega_{s}$ & $E^{(is)}=2(1-\eta_{ion})\eta_{ion}P^{(is)}_{2}+P^{(is)}_{loss}/(d+1)$\\
Initializing qubit f 　　　　&
$P^{(if)}_{2}=\Omega^{2}_{f}/2\textbf{B}^{2} $  &
$P^{(if)}_{loss}=\pi/\tau \Omega_{f}$ & $E^{(if)}=2P^{(if)}_{2}+P^{(if)}_{loss}/(d+1)$\\
Entangling qubit s and qubit f 　  &
$P^{(en)}_{2}=\Omega^{2}_{f}/8\textbf{B}^{2} $ &
$P^{(en)}_{loss}=5\pi/4\tau \Omega_{f}+ \pi/2\tau \Omega_{s}$& $E^{(en)}=P^{(en)}_{2}+P^{(en)}_{loss}/(d+1)$\\

Swapping 　　　& $P^{(sw)}_{2}=\Omega^{2}_{s}/8\textbf{B}^{2}$ &
$P^{(sw)}_{loss}=7\pi/4\tau \Omega_{s}$ & $E^{(sw)}=P^{(sw)}_{2}+P^{(sw)}_{loss}/(d+1)$ \\[0.5ex] \hline \hline
\end{tabular}
\end{center}
\end{table*}

Based on the above specific physical system, we now analyze the error sources
and determine the optimal Rabi frequencies for transitions in our
scheme. The main errors in our scheme arise in the procedures
involving Rydberg blockade, including the initializations of qubit s
(s=1,2) and qubit f, two-qubit phase gate for creating entanglement
between qubit s and qubit f, and two-qubit phase gate for entanglement
swapping. One can estimate the errors by adding the contributions
from the physically distinct processes of spontaneous emission from
Rydberg states and imperfect blockade errors, which are due to the
double excitation of Rydberg states. Using the techniques
developed in \cite{31}, we calculate the double excitation
probability $P_{2}$ and spontaneous emission probability $P_{loss}$
for the above four different procedures (as shown in Table I ).

It should be noted that in our system the cooperative
spontaneous emission dominates the decay processes from
Rydberg states. Thus the atoms in the Rydberg states are
most likely to decay to the reservoir state (with a
probability of order $d/(d+1)$, where $d$ is the optical
depth of the ensemble). If so, after spontaneous emission
the state of the ensemble will be outside the subspace
spanned by qubit $s$ and qubit $f$, and thereby can be
eliminated by the following post-selection measurement of
our repeater scheme. Hence, all the spontaneous emission
error terms are suppressed by a factor $1/(d+1)$. The
double excitation in the procedure of initializing qubit s
results in two ions in the selective ionization measurement
in the swapping, and thus induces an error with a
probability of
$2(1-\eta_{ion})\eta_{ion}P^{(is)}_{2}\approx0.1P^{(is)}_{2}$.
 A double excitation in the procedure of initializing qubit $f$ leads
to a two-photon emission into the fiber, which gives an error as
large as $2P^{(if)}_{2}$. Double excitations which occur in the two-qubit
gates for entanglement creation or entanglement swapping will cause
the final state to be separable. They introduce errors with
probabilities of $P_{2}^{(en)}$ and $P_{2}^{(sw)}$, respectively.
For clarity, we show all the errors corresponding to the
different procedures in Table I. The error in the entanglement
creation $E^{(c)}$ can be written as
\begin{equation}
\begin{split}
E^{(c)}&=E^{(is)}+E^{(if)}+E^{(en)}\\
&=\frac{0.1\Omega_{s}^{2}}{2\textbf{B}^2}
+\frac{3\pi}{2\tau\Omega_{s}(d+1)}+\frac{9\Omega_{f}^{2}}{8\textbf{B}^2}+\frac{9\pi}{4\tau\Omega_{f}(d+1)}.
\end{split}
\end{equation}
All the above errors result in a separable component
$\rho_{sep}$ in the created state with the probability
$E^{(c)}$, where the specific form of  $\rho_{sep}$ is not
important to our discussion. Hence, the density matrix of
each link after the entanglement creation reads
\begin{equation}
\rho_{0}\approx|\psi\rangle\langle\psi|+E^{(c)}\rho_{sep},
\end{equation}
where $|\psi\rangle$ is the desired entangled state and
$\rho_{sep}$ is the error term. The error term will be
amplified by subsequent swapping operations. Taking into
account the error in the entanglement swapping $E^{(sw)}$,
after the $n$-th swapping operation the density matrix of
final state is
\begin{equation}
\rho_{n}\approx|\psi\rangle\langle\psi|+[2^{n}E^{(c)}+(2^{n}-1)E^{(sw)}]\rho_{sep}.
\end{equation}

Note that for ultra-cold atoms trapped in an optical
lattice, the lifetime of a Rydberg state with $n=80$ is
about $500\mu s$. The optical depth $d=N\lambda^{2}/A$ in
such a $7\times7\times15$ optical lattice is around 10,
where $A$ is the cross section of the ensemble. Assuming a
nesting level $n=4$ (corresponding to $2^4=16$ links), one
can straightforwardly derive the optimal Rabi frequencies
$\Omega_{fopt}/2\pi=0.415$MHz and
$\Omega_{sopt}/2\pi=0.209$MHz, which minimize the error
term in Eq.(12), resulting in the fidelity of the final
entangled state $F\approx0.977$.

Now we use the parameters described above to estimate $p$, i.e., the
success probability for an ensemble to emit a photon. Suppose one
can collect the photon emitted in a direction within 0.3 rad off the
axis of the ensemble as in Ref. \cite{32}. Based on Eq.(3), we can predict that the photon is emitted into the collectable
area with more
than 93\% probability. Taking spontaneous emission in the preparation of
ensemble-photon entanglement state into account, we estimate $p\approx0.9$. The performance of a repeater with the above configuration is shown as
curve D in Fig.4.

Note that our scheme is quite robust to a reduction of $p$,
resulting in a significant flexibility of the requirements for the
atomic ensemble. For example, instead of the optical lattice, we could use
an atomic sample where 200 atoms are randomly positioned within a $6
\mu m$ diameter sphere, resulting in a moderate optical depth
$d\approx1$. Using the same method as above, we can derive that now
$p\approx0.2$ with the maximum fidelity of the final entangled state
$F\approx0.927$. As shown by curve C in Fig.4, the performance of
our scheme with this configuration still outperforms the DLCZ-type
repeater by at least one order of magnitude.

\subsection{Additional speed-up via temporal multiplexing}

As seen in the previous subsections, the creation of entanglement
between neighboring nodes A and B is heralded on the outcome of
photon detections at a middle station. To benefit from a nested
repeater, the entanglement swapping operations can only be performed
once one knows the relevant measurement outcomes. This requires a
communication time of order $L_{0}/c$. If every node consists of a
multiqubit register, and the entanglement creation in the register
can be triggered $m$ times in every communication time interval
$L_{0}/c$, one can decrease the average time for entanglement
creation $T_{link}$ by a factor of order $m$ \cite{15}. We here
propose a realization of the same basic idea for quantum repeaters
based on Rydberg blockade coupled ensembles.

As said before, the collective encoding strategy provides a
promising way to realize a multiqubit register. For instance,
Ref. \cite{33} proposed to use a single holmium ensemble to
realize a 60-qubit register via collective encoding. Note that in such a 60-qubit register every qubit can
be separately addressed and two-qubit gates between any two qubits are
achievable \cite{33}. Suppose we use such a 60-qubit register as
the quantum memory in each node. Take ensemble B as an example, one
of the 60 qubits in ensemble B is used to emit single photons, and
other 58 qubits are equally divided into two  groups (``red'' and
``blue'') as stationary qubits. Using the same procedure shown in
Fig.2, the qubits in the ``red'' and ``blue'' groups are alternately
entangled with corresponding emitted single photons, which are sent
toward ensembles A or C, respectively. If there are two detections
in the central station located between A and B for the $k$-th qubit
for example, then we know that these qubits are entangled. Running
the same scheme for the qubits in the other group, there may be
similar detections between B and C locations associated to the $l$-th
qubits. One then performs the Bell state analysis for entanglement
swapping by applying a two-qubit gate on the $k$-th and $l$-th qubits, thus creating entanglement between ensembles A and C.
Using such a holmium ensemble register could
increase the entanglement distribution rate of the studied scheme by up to a factor of 29.

\section{Conclusion}

We have shown that Rydberg blockade coupled ensembles are
very promising systems for the implementation of quantum
repeaters. Compared with DLCZ-type repeaters, our scheme
improves the entanglement distribution rate by several
orders of magnitude. One reason is that for Rydberg
blockade coupled atomic ensembles the entanglement swapping
operations are performed almost deterministically, in
contrast to success probabilities below 0.5 per swapping
for DLCZ-type repeaters. Another reason is that the
blockade mechanism suppresses multiple emissions from
individual ensembles so that our scheme does not need to
work with a very low emission probability. Compared with
repeaters based on trapped ions, both the entanglement
fidelity and the distribution rate of our scheme are
comparable.  This is because a Rydberg blockade coupled
atomic ensemble behaves as one superatom with a two-level
structure. However, by using an ensemble based scheme we
avoid the requirements of a high-finesse cavity, and of
addressing and transporting single ions. Moreover, our
scheme is amenable to temporal multiplexing, which could
further improve the performance.

{\it Acknowledgments.} We thank A. Lvovsky and A. MacRae for useful discussions.

{\it Note added.} After completion of this work we became aware of a very recent similar proposal \cite{zhao}.

\end{document}